\documentclass[aps,twocolumn,amssymb,amsmath,superscriptaddress]{revtex4-1}

\usepackage{bm}
\usepackage{color}
\usepackage{graphicx}
\usepackage{dcolumn}
\usepackage{graphicx}
\usepackage[colorlinks=true, letterpaper=true, pdfstartview=FitV, linkcolor=blue, citecolor=blue, urlcolor=blue]{hyperref}
\usepackage[normalem]{ulem}
\usepackage{amsmath}
\usepackage{epstopdf}

\setcitestyle{square}

\makeatletter
\renewcommand\@biblabel[1]{#1.}
\makeatother

\begin{document}

\title{Tunable magneto-optical effect, anomalous Hall effect and anomalous Nernst effect in two-dimensional room-temperature ferromagnet $1T$-CrTe$_2$}

\author{Xiuxian Yang}
\affiliation{Key Laboratory of Advanced Optoelectronic Quantum Architecture and Measurement, Ministry of Education, School of Physics, Beijing Institute of Technology, Beijing 100081, China}
\affiliation{Kunming Institute of Physics, Kunming 650223, China}

\author{Xiaodong Zhou}
\affiliation{Key Laboratory of Advanced Optoelectronic Quantum Architecture and Measurement, Ministry of Education, School of Physics, Beijing Institute of Technology, Beijing 100081, China}

\author{Wanxiang Feng}
\email{wxfeng@bit.edu.cn}
\affiliation{Key Laboratory of Advanced Optoelectronic Quantum Architecture and Measurement, Ministry of Education, School of Physics, Beijing Institute of Technology, Beijing 100081, China}

\author{Yugui Yao}
\affiliation{Key Laboratory of Advanced Optoelectronic Quantum Architecture and Measurement, Ministry of Education, School of Physics, Beijing Institute of Technology, Beijing 100081, China}

\date{\today}

\begin{abstract}
Utilizing the first-principles density functional theory calculations together with group theory analyses, we systematically investigate the spin order-dependent magneto-optical effect (MOE), anomalous Hall effect (AHE), and anomalous Nernst effect (ANE) in a recently discovered two-dimensional room-temperature ferromagnet $1T$-CrTe$_2$.  We find that the spin prefers an in-plane direction by the magnetocrystalline anisotropy energy calculations.  The MOE, AHE, and ANE display a period of $2\pi/3$ when the spin rotates within the atomic plane, and they are forbidden if there exists a mirror plane perpendicular to the spin direction.  By reorienting the spin from in-plane to out-of-plane direction, the MOE, AHE, and ANE are enhanced by around one order of magnitude.  Moreover, we establish the layer-dependent magnetic properties for multilayer $1T$-CrTe$_2$ and predict antiferromagnetism and ferromagnetism for bilayer and trilayer $1T$-CrTe$_2$, respectively.  The MOE, AHE, and ANE are prohibited in antiferromagnetic bilayer $1T$-CrTe$_2$ due to the existence of the spacetime inversion symmetry, whereas all of them are activated in ferromagnetic trilayer $1T$-CrTe$_2$ and the MOE is significantly enhanced compared to monolayer $1T$-CrTe$_2$.  Our results show that the magneto-optical and anomalous transports proprieties of $1T$-CrTe$_2$ can be effectively modulated by altering spin direction and layer number.
\end{abstract}

\maketitle

\section{Introduction}\label{intro}
Although two-dimensional (2D) materials have been explored for more than a decade, the magnetic order rarely survives in atomically thin films due to thermal fluctuations~\cite{Gong2019}.  The realization of 2D magnets is a big challenge~\cite{Mermin1966} and has attracted extensive attention~\cite{Hellman2017,Burch2018}.  The 2D magnetic van der Waals (vdW) materials are especially expected to open up a wide range of possibilities for spintronics~\cite{Sachs2013,Park2016,Zhong2017}.  Thanks to the improvement of theoretical methods and experimental capabilities, more and more 2D magnetic vdW materials have been discovered, which indicates that the field of 2D magnets is advancing rapidly~\cite{Samarth2017}.  In recent years, for example, tens of 2D vdW materials with stable magnetic orders have been observed in layered FePS$_3$~\cite{Lee2016,Wang2016}, Cr$_2$Ge$_2$Te$_6$~\cite{Gong2017}, Cr$X_3$($X$=I, Br, Cl)~\cite{Huang2017,Klein2018,Jiang2018,Huang2018,Jiang2018_2,Wang2018,Sivadas2018,Kim2019,Klein2019,Kim2019_2}, Fe$_3$GeTe$_2$~\cite{Deng2018,Fei2018,Xu2020}, $MX_2$($M$=V, Mn; $X$=Se, Te)~\cite{Bonilla2018,Li2018,OHara2018}, MnSn~\cite{Yuan2020}, PtSe$_2$~\cite{Avsar2019}, and CrTe$_2$~\cite{Freitas2015,Sun2020,Purbawati2020}.

Aimed at the applications of 2D spintronics, detecting spontaneous magnetization is the primary step.  The standard techniques, such as superconducting quantum interference device (SQUID) magnetometer and neutron scattering, are challenging to use for 2D magnetic vdW materials~\cite{Burch2018,Hellman2017}.  Instead, the magneto-optical effects (MOE), represented by the Kerr~\cite{Kerr1877} and Faraday~\cite{Faraday1846} effects, are considered to be a powerful and non-contact (non-destructive) probe of magnetism in 2D materials~\cite{Gong2017,Huang2017}.  The magneto-optical Kerr and Faraday effects are defined as the rotation of the polarization planes of reflected and transmitted light beams when a linearly polarized light hits the magnetic materials~\cite{Antonov2014}.  In condensed matter physics, the MOE and the anomalous Hall effect (AHE)~\cite{Nagaosa2010}, where the latter is characterized by a transverse voltage generated by a longitudinal charge current in the absence of external magnetic fields, are two fundamental phenomena that usually coexist in ferromagnets and antiferromagnets.  There are two distinct contributions to the AHE, that is the extrinsic AHE (i.e., side jump and skew scattering) depending on the scattering of electron off impurities or due to disorder, and the intrinsic AHE expressed in term of Berry curvatures in a perfect crystal~\cite{Nagaosa2010}.  According to the Kubo formula~\cite{Kubo1957,Wang1974}, the intrinsic anomalous Hall conductivity (AHC) can be straightforwardly extended to the optical Hall conductivity, which is intimately related to the magneto-optical Kerr and Faraday effects~\cite{Zhou2019}.  Because of the inherent relationship between the intrinsic AHE and MOE, they are often studied together. Moreover, the transverse charge current can also be generated by a longitudinal temperature gradient, called anomalous Nernst effect (ANE)~\cite{Nernst1887}, which has attracted enormous interest mainly due to its promising applications on the thermoelectric aspects.  The giant ANE has been recently discovered in chiral magnets~\cite{Hanasaki2008,Ikhlas2017} and topological semimetals~\cite{Liang2017,Wuttke2019}.

\begin{figure*}
	\includegraphics[width=1.5\columnwidth]{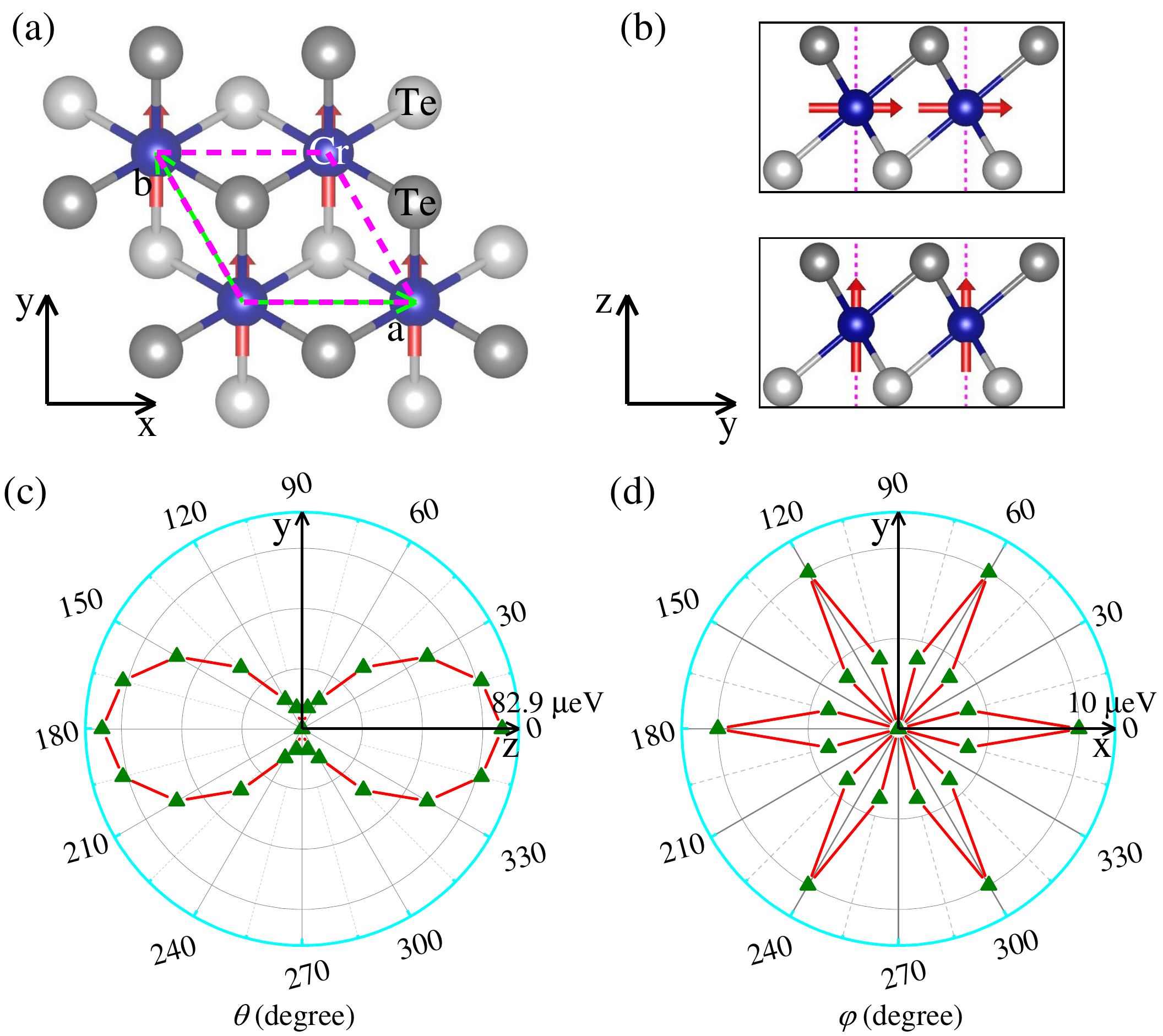}
	\caption{(Color online) (a,b) Top and side views of monolayer $1T$-CrTe$_2$.  The blue spheres represent Cr atoms, whereas dark-gray and silver-white spheres represent Te atoms in the upper and lower sublayers.  The pink dashed lines draw up the 2D primitive cell, and the red arrows indicate the directions of spin magnetic moments.  The top and bottom panels in (b) present the spin directions along the $y$- and $z$-axis, respectively.  (c,d) The magnetocrystalline anisotropy energy of monolayer $1T$-CrTe$_2$ by rotating the spin magnetic moment within the $yz$ and $xy$ planes.  The spin along the $y$-axis is set to be the reference state.}
	\label{fig:crystal}
\end{figure*}

The catalog of 2D magnetic vdW materials is rich; however, the ferromagnetic candidates with high Curie temperatures ($T_C$) are still limited, hindering enormously the development of 2D spintronics.  Fortunately, a 2D dichalcogenide with the $1T$ polytype, $1T$-CrTe$_2$ [see Figs.~\ref{fig:crystal}(a) and~\ref{fig:crystal}(b)], has been recently synthesized with an exceptionally high $T_C$ ($>300$ K)~\cite{Freitas2015,Sun2020,Purbawati2020}.  In this work, based on the first-principles density functional theory calculations and group theory analyses, we systematically investigate the electronic, magnetic, magneto-optical, and anomalous charge and thermoelectric transports properties of monolayer and multilayer $1T$-CrTe$_2$ (hereafter, we use CrTe$_2$ for simplification).  We find that monolayer CrTe$_2$ is a ferromagnetic metal with the in-plane  magnetization direction.  By calculating magnetocrystalline anisotropy energy (MAE), the magnetization direction is finely identified along the $y$-axis [see Fig.~\ref{fig:crystal}(a)], and the maximal value of MAE between in-plane and out-of-plane magnetization directions reaches to 82.9 $\mu$eV/cell, which is much smaller than that of famous 2D ferromagnets CrI$_3$ (1.37 meV/cell)~\cite{Zhang2015} and Fe$_3$GeTe$_2$ (2.76 meV/cell)~\cite{Zhuang2016}.  It indicates that the spin direction of monolayer CrTe$_2$ can be easily tuned by an external magnetic field.  The MOE, AHE, and ANE display a period of $2\pi/3$ by rotating the spin within the $xy$ plane, and they disappear if there exists a mirror plane perpendicular to the spin direction.  We then show that changing the spin from in-plane to out-of-plane direction can enhance the MOE, AHE, and ANE by around one order of magnitude.  Additionally, the layer-dependent magnetic properties for multilayer CrTe$_2$ are studied, and antiferromagnetism and ferromagnetism for bilayer and trilayer CrTe$_2$ are predicted, respectively.  For antiferromagnetic bilayer CrTe$_2$, the MOE, AHE, and ANE are fully suppressed due to the existence of the spacetime inversion symmetry $\mathcal{T} \mathcal{P}$ ($\mathcal{T}$ and $\mathcal{P}$ are time-reversal and spatial inversion operations, respectively).  In contrast, the MOE, AHE, and ANE are activated in ferromagnetic trilayer CrTe$_2$, and the MOE is significantly enhanced compared to that of monolayer CrTe$_2$.  Our results show that the magneto-optical and anomalous transports proprieties of 2D CrTe$_2$ are tunable by altering the magnetization direction and the number of layers.

\section{Methodology}\label{method}
The first-principles calculations were performed by Vienna \textit{ab initio} simulation package (\textsc{vasp})~\cite{Kresse1993,Kresse1996} within the framework of density functional theory.  The projector augmented wave method (PAW)~\cite{Blochl1994} was employed to model the ion cores and the exchange-correlation functional of generalized gradient approximation (GGA) with the Perdew-Burke-Ernzerhof parameterization (PBE)~\cite{Perdew1996} to simulate the valence electrons.  Spin-orbit coupling was included in the calculations for the MAE, MOE, AHE, and ANE.  The plane-wave cutoff energy was set to be 500 eV.  The structures were relaxed until the maximum force on each atom is less than $0.0001$ eV/{\AA} and the energy convergence criterion is $10^{-7}$ eV.  The Brillouin zone integration was carried out by $16\times16\times1$ $k$-points sampling.  A vacuum layer with the thickness of at least 15 {\AA}  was used to avoid the interactions between adjacent layers and the vdW correction was adopted by DFT-D2 method in multilayer structures.  The optical conductivity, AHC, and ANC are scaled by a factor of $Z/d_{0}$ to exclude the vacuum region, where $Z$ is the cell length normal to the atomic plane and $d_{0}$ = 6.23 \AA, 12.46 \AA, and 18.69 \AA\ are the effective thicknesses of monolayer, bilayer, and trilayer CrTe$_2$, respectively.  Since the $d$ orbitals of  Cr atom are not fully filled, the LDA+U method~\cite{Anisimov1991,Dudarev1998} was used to account for the Coulomb correlation with U = 2.0 eV~\cite{Sui2017}.

To obtain the MOE, such as the Kerr and Faraday spectra, the optical conductivity should be primarily calculated.  Here, we constructed the maximally localized Wannier functions (MLWFs) in a non-self-consistent process by projecting onto $s$, $p$, and $d$ orbitals of Cr atom as well as onto $s$ and $p$ orbitals of Te atom, using a uniform $k$-mesh of $16\times16\times1$ points in conjunction with the \textsc{wannier90} package~\cite{Mostofi2008}.  The optical conductivity was then calculated by integrating the dipole matrix elements (under the MLWFs basis) over the entire Brillouin zone using a very dense $k$-points of $300\times300\times1$.  The absorptive parts of optical conductivity are given by~\cite{Kubo1957,Wang1974,Callaway2013},
\begin{eqnarray}
\sigma^1_{xx}(\omega)&=&\frac{\lambda}{\omega}\sum_{\textbf{k},jj'}[|\Pi^+_{jj'}|^2+|\Pi^-_{jj'}|^2]\delta(\omega-\omega_{jj'}), \label{eq:Optical_xx1}\\
\sigma^2_{xy}(\omega)&=&\frac{\lambda}{\omega}\sum_{\textbf{k},jj'}[|\Pi^+_{jj'}|^2-|\Pi^-_{jj'}|^2]\delta(\omega-\omega_{jj'}), \label{eq:Optical_xy2}
\end{eqnarray}
where the superscripts 1 and 2 indicate the real and imaginary parts, $\lambda=\frac{\pi e^2}{2\hbar m^2V}$ is a material specific constant ($e$ and $m$ are the charge and mass of an electron, $\hbar$ is reduced Planck constant, and $V$ is volume of unit cell), $j$ and $j'$ denote occupied and unoccupied states at the same $k$-point, $\Pi^\pm_{jj'}$ are the dipole matrix elements relevant to right-circularly ($+$) and left-circularly ($-$) polarized lights, $\hbar\omega$ is the photon energy, $\hbar\omega_{jj'}$ is the energy difference between $j$ and $j'$ states.  Utilizing the Kramers-Kronig transformation, the dispersive parts can be obtained as,
\begin{eqnarray}
\sigma^2_{xx}(\omega)&=&-\frac{2}{\pi}\mathcal{P}\int^\infty_0\frac{\sigma_{xx}^1(\omega')}{\omega'^2-\omega^2}d\omega', \label{eq:Optical_xx2} \\
\sigma^1_{xy}(\omega)&=&\frac{2}{\pi}\mathcal{P}\int^\infty_0\frac{\omega\sigma_{xy}^2(\omega')}{\omega'^2-\omega^2}d\omega', \label{eq:Optical_xy1}
\end{eqnarray}
where $\mathcal{P}$ is the principal integral.  

The Kerr effect is characterized by the rotation angle ($\theta_\textnormal{K}$) and ellipticity ($\varepsilon_\textnormal{K}$), which are usually combined into the complex Kerr angle,
\begin{equation}\label{eq:Kerr}
\phi_\textnormal{K}=\theta_\textnormal{K}+i\varepsilon_\textnormal{K}=i\frac{2\omega d}{c}\frac{\sigma_{xy}}{\sigma_{xx}^s},
\end{equation}
where $c$ is the speed of light in vacuum, $d$ is the thin-film thickness, and $\sigma_{xx}^s$ is the optical conductivity of a nonmagnetic substrate.  Similarly, the complex Faraday angle is given by,
\begin{equation}\label{eq:Faraday}
\phi_\textnormal{F}=\theta_\textnormal{F}+i\varepsilon_\textnormal{F}=i\frac{\omega d}{2c}(n_+-n_-),
\end{equation}
where $n_\pm^2=1+\frac{4\pi i}{\omega}(\sigma_{xx}\pm i\sigma_{xy})$ are eigenvalues of dielectric tensor.  By considering the fact that $|\frac{4\pi i}{\omega}(\sigma_{xx}\pm i\sigma_{xy})|\ll1$, the complex Faraday angle can be approximately written as,
\begin{equation}\label{eq:Faraday2}
\phi_\textnormal{F}=\theta_\textnormal{F}+i\varepsilon_\textnormal{F}\simeq -\frac{2\pi d}{c}\sigma_{xy},
\end{equation}
From Eqs.~\eqref{eq:Kerr} and~\eqref{eq:Faraday2}, one can see that the off-diagonal elements of optical conductivity ($\sigma_{xy}$), also known as the optical Hall conductivity, is determinative to both Kerr and Faraday effects.  It should be mentioned here that Eqs.~\eqref{eq:Kerr}--\eqref{eq:Faraday2} are the expressions for 2D systems with a polar geometry~\cite{Feng2017}, that is, the incident light propagates along the $-z$ direction.

Physically speaking, the MOE is closely related to the AHE.  For example, the dc limit of the real part of the off-diagonal element of optical conductivity, i.e., $\sigma^1_{xy} (\omega\rightarrow0)$, is nothing but the intrinsic AHC, which can also be calculated from the Berry-phase formula~\cite{Yao2004},
\begin{equation}\label{eq:AHC}
\sigma^\textnormal{A}_{xy}=-\frac{e^2}{\hbar V}\sum_{n,\textbf{k}}f_{n\textbf{k}}\Omega^n_{xy}(\textbf{k}),
\end{equation}
where $n$, $\textbf{k}$, and $f_{n\textbf{k}}$ are band index, crystal momentum, and Fermi-Dirac distribution function, respectively.  $\Omega^n_{xy}(\textbf{k}$) is the band-resolved Berry curvature, given by,
\begin{equation}\label{eq:Berry}
\Omega^n_{xy}(\textbf{k})=-\sum_{n'\neq n}\frac{2 {\rm Im}[\langle\psi_{n\textbf{k}}|\hat{v}_x|\langle\psi_{n'\textbf{k}}\rangle\langle\psi_{n'\textbf{k}}|\hat{v}_y|\langle\psi_{n\textbf{k}}\rangle]}{(\varepsilon_{n\textbf{k}}-\varepsilon_{n'\textbf{k}})^2}
\end{equation}
where $\hat{v}_{x,y}$ is the velocity operator along the $x$ or $y$ direction, $\psi_{n\textbf{k}}$ and $\varepsilon_{n\textbf{k}}$ are the eigenvector and eigenvalue at band index $n$ and crystal momentum $\textbf{k}$, respectively.  The intrinsic anomalous Nernst conductivity (ANC) can be written as~\cite{Xiao2006,Zhou2020}
\begin{eqnarray}\label{eq:ANC}
\alpha^\textnormal{A}_{xy}&=&\frac{e}{\hbar TV}\sum_{n,\textbf{k}}\Omega_{xy}^n(\textbf{k})\times[(\varepsilon_{n\textbf{k}}-\mu)f_{n\textbf{k}} \nonumber \\
&&+k_BT {\rm ln}(1+e^{-(\varepsilon_{n\textbf{k}}-\mu)/k_BT})]
\end{eqnarray}
where $T$, $\mu$, and $k_B$ is temperature, chemical potential, and Boltzmann constant, respectively.  Thus, the ANC can be related to the AHC by the Mott formula~\cite{Xiao2006}.

\begin{figure}
	\includegraphics[width=\columnwidth]{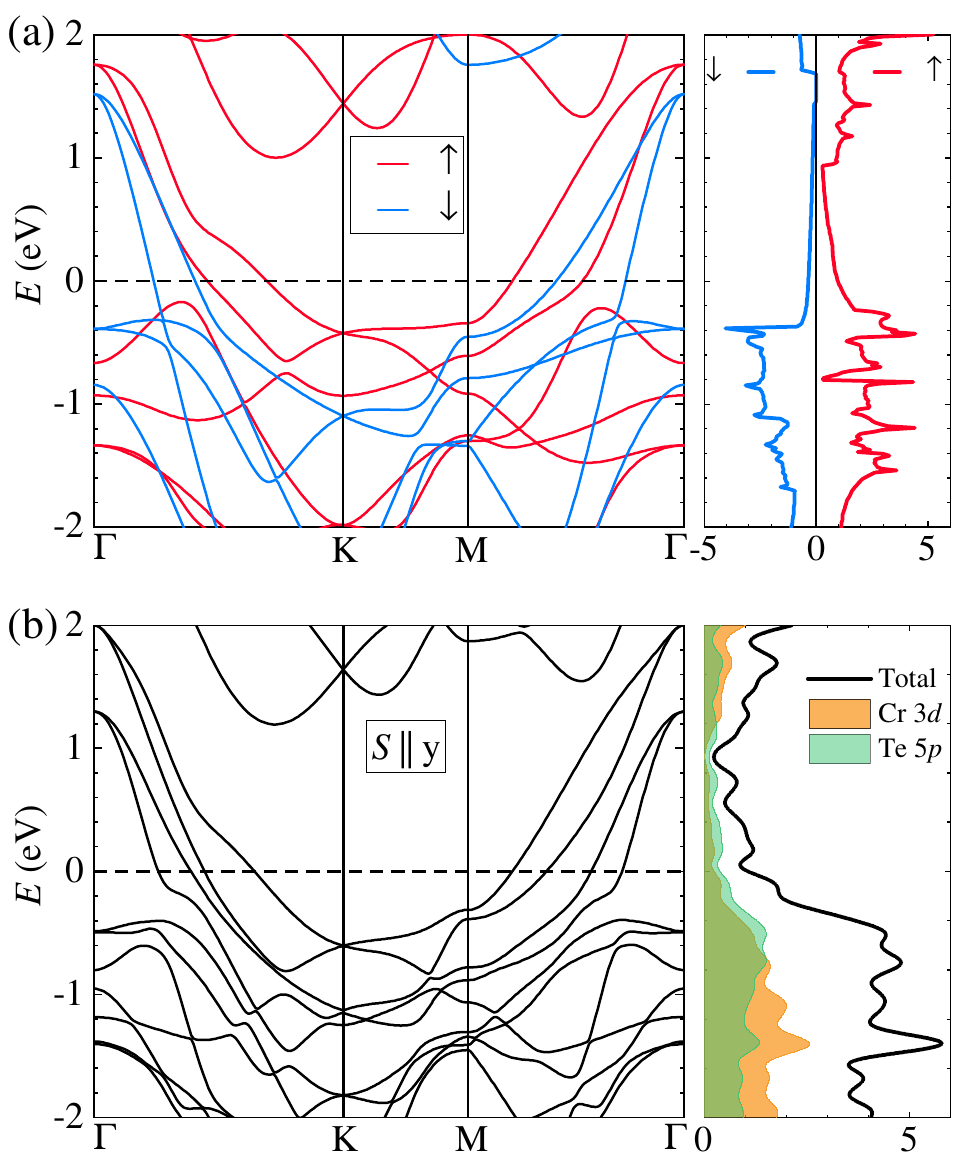}
	\caption{(Color online) (a) The spin-polarized band structure and density of states (in the unit of states/eV/cell) for monolayer $1T$-CrTe$_2$.  (b) The relativistic band structure and orbital-decomposed density of states (in the unit of states/eV/cell) for monolayer $1T$-CrTe$_2$ when the spin is along the $y$-axis.}
	\label{fig:band}
\end{figure}

\section{Results and discussion}\label{results}
In this section, we successively present the results of monolayers and multilayer CrTe$_2$.  The magnetic ground states are first established by calculating the MAE.  The corresponding electronic band structures are explicitly calculated.  The magnetic group theory is then used to determine the nonzero elements of optical conductivity, which is the critical ingredient to evaluate the magneto-optical Kerr and Faraday spectra.  Finally, the anomalous Hall and anomalous Nernst conductivities are evaluated by using the Berry-phase formulas.  The dependence of the MOE, AHE, and ANE on the magnetization direction and layer number will be detailedly discussed.

\subsection{Monolayer CrTe$_2$}
\subsubsection{Crystal, magnetic, and electronic structures}
The top and side views of monolayer CrTe$_2$ (space group P$\overline{3}$m$1$, No. $164$) are depicted in Figs.~\ref{fig:crystal}(a) and~\ref{fig:crystal}(b).  Each primitive cell contains one Chromium (Cr) atom and two Tellurium (Te) atoms, forming a sandwich structure Te-Cr-Te.  The optimized lattice constant of monolayer CrTe$_2$ is $a = 3.722$ {\AA}.

To confirm the magnetic ground state, we compared total energies among the nonmagnetic, antiferromagnetic, and ferromagnetic states using a supercell of $2\times2\times1$, and the results show that ferromagnetic state is more stable than nonmagnetic and antiferromagnetic states by $2.90$ eV and $55.32$ meV, respectively.  Additionally, the magnetic ground state can be determined via the super-exchange mechanism~\cite{Anderson1950,Goodenough1958,Kanamori1959}.  The magnetic exchange interactions depend on the filling of the $d$ orbitals of the cations and on the angle formed by the chemical bonds connecting the ligand and magnetic atoms, and particularly, when the angle equals to $90^\circ$ the ferromagnetic interactions are optimal.  In the case of CrTe$_2$, the bond angle between Cr-Te-Cr is $87^\circ$, which accounts for ferromagnetic interactions.  Furthermore, the MAE, defined as $\textnormal{MAE}(\theta,\varphi)=E(\theta,\varphi)-E(\theta=90^\circ,\varphi=90^\circ)$ [here, $E(\theta,\varphi)$ is the total energy when the spin magnetic moment ($S$) orients to the polar angle $\theta$ and azimuthal angle $\varphi$], is computed by rotating the spin magnetic moment on the $xy$ and $yz$ planes, respectively, as shown in Figs.~\ref{fig:crystal}(c) and~\ref{fig:crystal}(d).  The positive values of MAE suggest a preferred magnetization along the $y$-axis ($\theta=90^\circ,\varphi=90^\circ$) rather than along other directions.  Figure~\ref{fig:crystal}(c) shows that the out-of-plane magnetization (along the $z$ axis) is not prior due to the positive MAE of $82.9$ $\mu$eV/cell.  Figure~\ref{fig:crystal}(d) further indicates a small in-plane magnetocrystalline anisotropy ($0$ $\mu$eV/cell $\leq$ MAE $\leq$ $10$ $\mu$eV/cell), in good agreement with experimental observation~\cite{Sun2020,Purbawati2020}.  Therefore, the magnetic ground state of the system is confirmed and the spin magnetic moment should be along the $y$-axis, i.e., $S(\theta=90^\circ,\varphi=90^\circ)$ or $S\parallel y$ [see top panel of Fig.~\ref{fig:crystal}(b)], which is consistent with previously theoretical calculation~\cite{Lv2015}.  Thus, the spin direction of CrTe$_2$ can be easily tuned by applying an external magnetic field as the maximal MAE of CrTe$_2$ ($82.9$ $\mu$eV/cell) is much smaller than that of Fe$_3$GeTe$_2$ ($2.76$ meV/cell)~\cite{Zhuang2016}, which has been realized experimentally.  This provides a technical basis for us to reorient the spin magnetic moment from in-plane to out-of-plane direction [see bottom panel of Fig.~\ref{fig:crystal}(b)].

\begin{table*}[htpb]\footnotesize
	\caption{The magnetic space group (MSG) and magnetic point group (MPG) of monolayer $1T$-CrTe$_2$ as a function of azimuthal ($\varphi$) and polar ($\theta$) angles when the spin rotates within the $xy$ ($\theta=\pi/2$, $0\leq\varphi \leq\pi$) and $yz$ ($0\leq\theta \leq\pi$, $\varphi=\pi/2$) planes.}
	\label{tab:group}
	\begin{ruledtabular}
		\begingroup
		\setlength{\tabcolsep}{4.5pt} % Default value: 6pt
		\renewcommand{\arraystretch}{1.5} % Default value: 1
		\begin{tabular}{lccccccccccccc}
			
			&$0^{\circ}$&$15^{\circ}$&$30^{\circ}$&$45^{\circ}$&$60^{\circ}$&$75^{\circ}$&$90^{\circ}$&$105^{\circ}$&$120^{\circ}$&$135^{\circ}$&$150^{\circ}$&$165^{\circ}$&$180^{\circ}$ \\
			
			\hline
			
			MSG($\varphi$) & $C2/m$ & $P\bar{1}$ & $C2'/m'$ & $P\bar{1}$ & $C2/m$ & $P\bar{1}$ & $C2'/m'$ & $P\bar{1}$ & $C2/m$ & $P\bar{1}$ & $C2'/m'$ & $P\bar{1}$ & $C2/m$ \\
			
			MPG($\varphi$) & $2/m$ & $\bar{1}$ & $2'/m'$ & $\bar{1}$ & $2/m$ & $\bar{1}$ & $2'/m'$ & $\bar{1}$ & $2/m$ & $\bar{1}$ & $2'/m'$ & $\bar{1}$  & $2/m$ \\

			MSG($\theta$) & $P\bar{3}m'1$ & $C2'/m'$ & $C2'/m'$ & $C2'/m'$ & $C2'/m'$ & $C2'/m'$ & $C2'/m'$ & $C2'/m'$ & $C2'/m'$ & $C2'/m'$ & $C2'/m'$ & $C2'/m'$ & $P\bar{3}m'1$ \\
			
			MPG($\theta$) & $\bar{3}1m'$ & $2'/m'$ & $2'/m'$ & $2'/m'$ & $2'/m'$ & $2'/m'$ & $2'/m'$ & $2'/m'$ & $2'/m'$ & $2'/m'$ & $2'/m'$ & $2'/m'$ & $\bar{3}1m'$
		\end{tabular}
		\endgroup
	\end{ruledtabular}
\end{table*}

We then discuss the electronic structures of monolayer CrTe$_2$.   Fig.~\ref{fig:band}(a) plots the spin-polarized band structures and density of states, in which the red and blue lines represent the spin-up ($\uparrow$) and spin-down ($\downarrow$) bands, respectively.  The spin-polarized band structures combined with density of states clearly show that monolayer CrTe$_2$ is a ferromagnetic metal.   After including spin-orbit coupling, the relativistic band structures and orbital-decomposed density of states with the magnetization $S\parallel y$ are illustrated in Fig.~\ref{fig:band}(b).  The band structure is good consistent with a recently theoretical calculation~\cite{Li2020}.  For the density of states, we only present the dominant components, i.e., the $3d$ orbitals of Cr atom (the orange pattern) and $5p$ orbitals of Te atoms (the green pattern), which have nearly equal contributions around the Fermi energy.

\subsubsection{Magnetic group theory}
The group theory is a powerful tool for identifying the nonvanishing elements of the optical Hall conductivity, which is the key factor in predicting the MOE.  Additionally, the AHE and ANE have the same symmetry requirements with the MOE due to their physical relations [refer to Eqs.~\eqref{eq:Optical_xx1}--\eqref{eq:ANC}].  Hence, we take the optical Hall conductivity as an example, and the results of symmetry analyses are applicable to the MOE, AHE, and ANE.  The magnetic space and point groups for monolayer CrTe$_2$ are calculated by using the \textsc{isotropy} software~\cite{isotropy}.  Table~\ref{tab:group} lists the results when the spin rotates within the $xy$ and $yz$ planes.   Since the optical Hall conductivity is translationally invariant, it is sufficient to restrict the analysis to magnetic point group.  Moreover, the vector-form notation of the optical Hall conductivity, given by $\bm{\sigma}(\omega)=[\sigma^x,\sigma^y,\sigma^z]=[\sigma_{yz},\sigma_{zx},\sigma_{xy}]$, is used for convenience as it can be regarded as a pseudovector, just like spin.  Thus, for a 2D system, there always has $\sigma^x=\sigma^y=0$ and only $\sigma^z$ is potentially nonzero.

Let us start with the situation that rotating the spin within the $xy$ plane.  The magnetic point group has a period of $\pi/3$: $2/m\rightarrow\bar{1}\rightarrow2'/m'\rightarrow\bar{1}\rightarrow2/m$, and three nonrepetitive elements are $2/m$, $\bar{1}$, and $2'/m'$.  First, the group $2/m$ (when $\varphi=n\pi/3$ with $n\in \mathbb{N}$) has a mirror plane that is parallel to the $z$-axis and is perpendicular to the spin direction.  Such a mirror operation reverses the sign of $\sigma^z$, and thus indicating $\sigma^z=0$.  It results in the vanishing optical Hall conductivity, i.e., $\bm{\sigma}(\omega)=[0,0,0]$.  On the other hand, all mirror symmetries are broken if $\varphi\neq n\pi/3$.  The group $2'/m'$ contains a combined symmetry $\mathcal{TM}$, where $\mathcal{T}$ is the time-reversal symmetry and $\mathcal{M}$ is a mirror plane that parallels to both the $z$-axis and spin direction.  Both $\mathcal{T}$ and $\mathcal{M}$ operations reverse the sign of $\sigma^z$, and hence $\sigma^z$ is even under $\mathcal{TM}$ symmetry.  It gives rise to the nonvanishing optical Hall conductivity, $\bm{\sigma}(\omega)=[0,0,\sigma^z]$.  Finally, for the group $\bar{1}=\{E,\mathcal{P}\}$, none of its elements (unit operation $E$ and spatial inversion $\mathcal{P}$) can affect $\sigma^z$, and hence the optical Hall conductivity are absolutely allowed.

We next turn to the case that the spin lies within the $yz$ plane.  The evolution of magnetic point group exhibits a period of $\pi$, and only two groups $\bar{3}1m'$ and $2'/m'$ are needed to analyze.  If $\theta=0$ or $\pi$, the group $\bar{3}1m'$ contains three $\mathcal{TM}$ symmetries with $\mathcal{M}\parallel z$.  The nonvanishing optical Hall conductivity can be expected since any one of the three $\mathcal{TM}$ symmetries affords $\sigma^z\neq0$.  Once the spin cants away from the $z$-axis ($\theta\neq0$) or from the $-z$-axis ($\theta\neq\pi$), the magnetic point group changes to be $2'/m'$, in which only one of the three $\mathcal{TM}$ symmetry leaves (here, $\mathcal{M}$ is just the $yz$ plane) but still ensures $\sigma^z\neq0$.  To summarize, the optical Hall conductivity is nonzero when the spin lies within the $yz$ plane, that is, $\bm{\sigma}(\omega)=[0,0,\sigma^z]$.

\begin{figure*}
	\includegraphics[width=2\columnwidth]{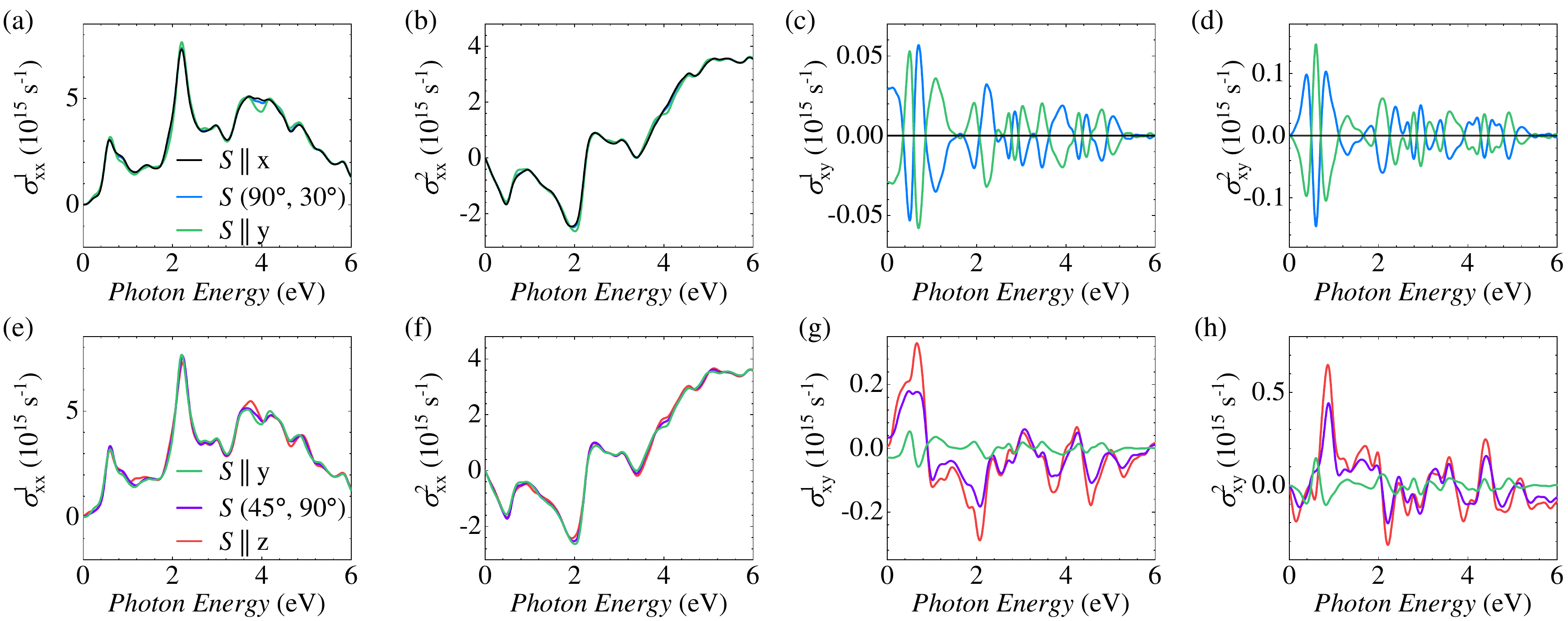}
	\caption{(Color online) The real diagonal (a,e), imaginary diagonal (b,f), real off-diagonal (c,g), and imaginary off-diagonal (d,h) elements of optical conductivity for monolayer $1T$-CrTe$_2$ with in-plane and out-of-plane magnetization, respectively.  For a better comparison, the curves when spin points along $y$-axis ($S\parallel y$) are replotted in (e-h).}
	\label{fig:optical}
\end{figure*}

\begin{figure*}
	\includegraphics[width=2\columnwidth]{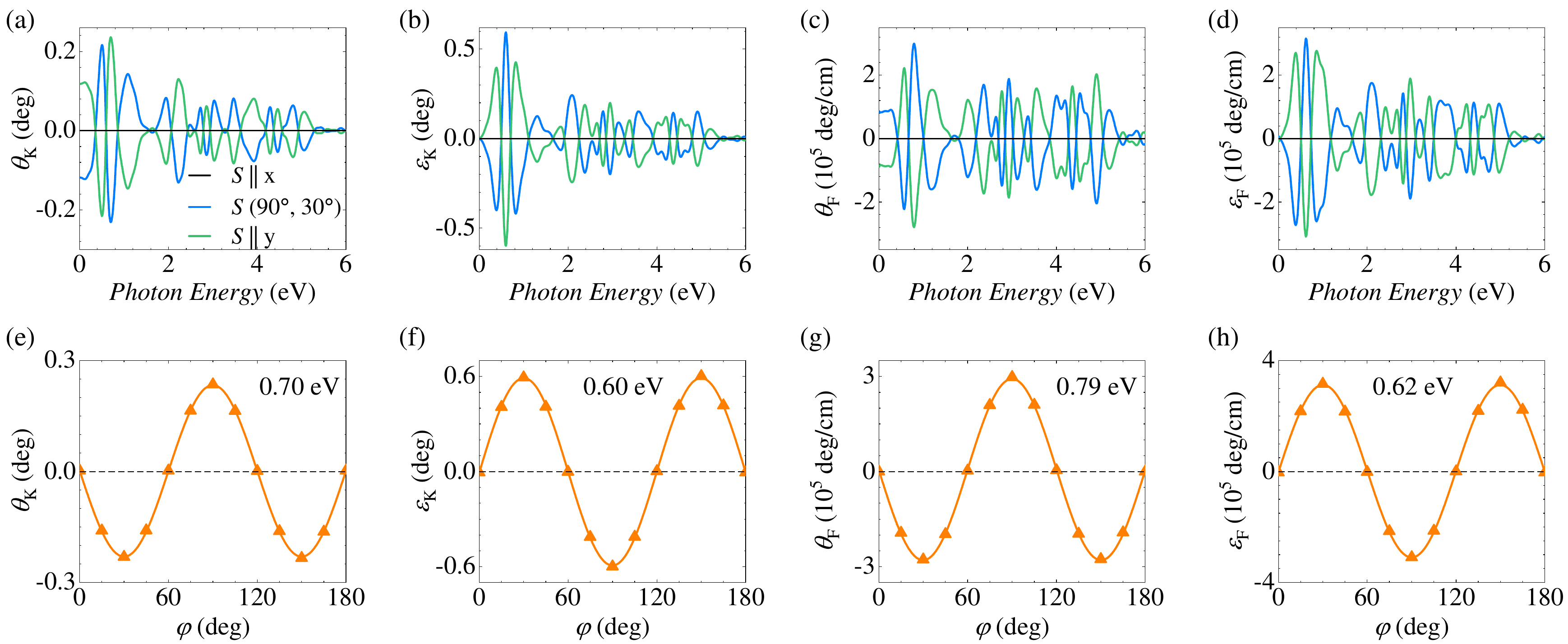}
	\caption{(Color online) The Kerr rotation angle $\theta_\textnormal{K}$ (a), Kerr ellipticity $\varepsilon_\textnormal{K}$ (b), Faraday rotation angle $\theta_\textnormal{F}$ (c), and Faraday ellipticity $\varepsilon_\textnormal{F}$ (d) for monolayer $1T$-CrTe$_2$ with in-plane magnetization.  (e-h) The Kerr and Faraday rotation angles and ellipticities as a function of azimuthal angle $\varphi$ at selected photon energies.}
	\label{fig:MOE1}
\end{figure*}

\begin{figure*}
	\includegraphics[width=2\columnwidth]{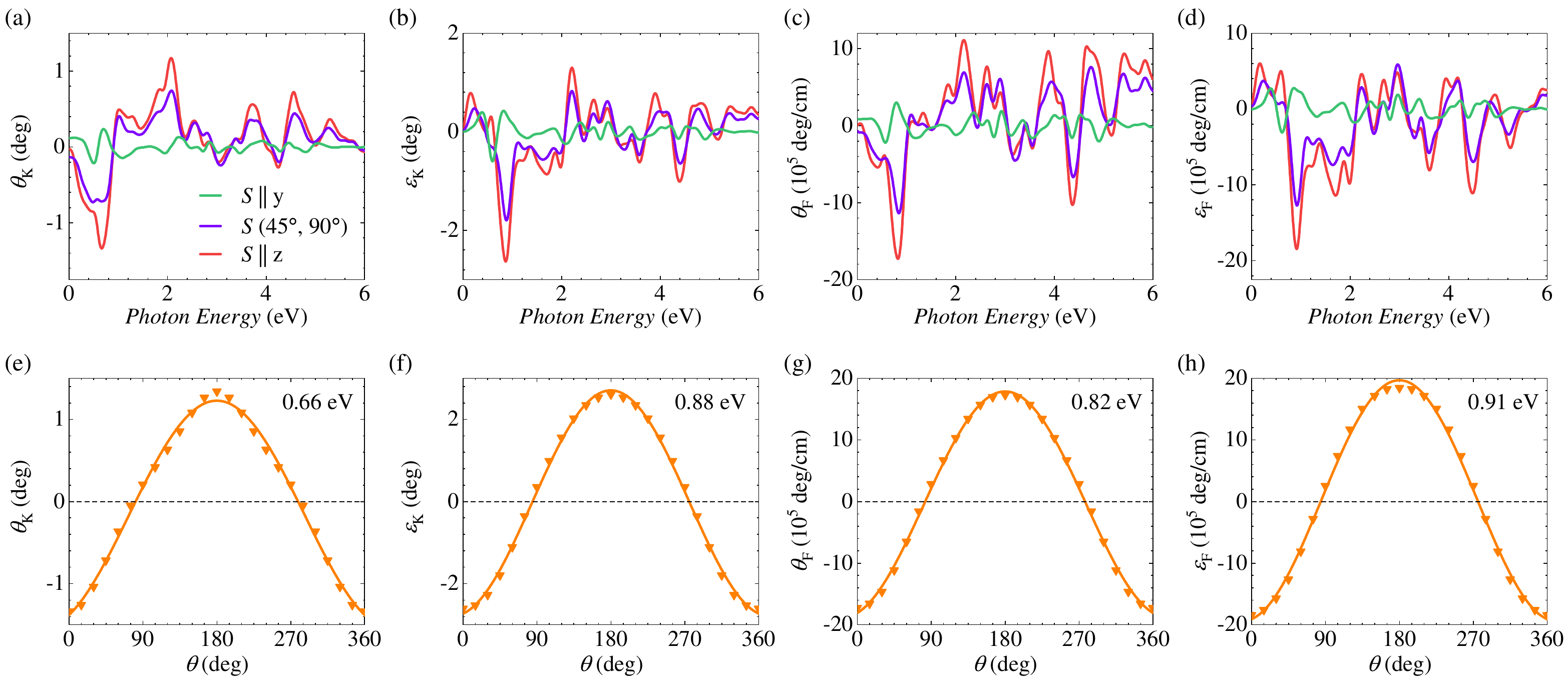}
	\caption{(Color online) The Kerr rotation angle $\theta_\textnormal{K}$ (a), Kerr ellipticity $\varepsilon_\textnormal{K}$ (b), Faraday rotation angle $\theta_\textnormal{F}$ (c), and Faraday ellipticity $\varepsilon_\textnormal{F}$ (d) for monolayer $1T$-CrTe$_2$ with out-of-plane magnetization.  For a better comparison, the curves for the spin pointing along the $y$-axis ($S\parallel y$) are also plotted in (a-d).  (e-h) The Kerr and Faraday rotation angles and ellipticities as a function of polar angle $\theta$ at selected photon energies.}
	\label{fig:MOE2}
\end{figure*}

\subsubsection{Optical and magneto-optical properties}

After obtaining the electronic structures and magnetic groups of monolayer CrTe$_2$ with different magnetization directions, we now focus on the optical conductivity, which is prerequisite to evaluate the MOE.

We first discuss the results of in-plane magnetization when $S(90^\circ, 0^\circ)$ ($S\parallel x$), $S(90^\circ, 30^\circ)$, and $S(90^\circ, 90^\circ)$ ($S\parallel y$), shown in Figs.~\ref{fig:optical}(a-d).  According to Eqs.~\eqref{eq:Optical_xx1} and~\eqref{eq:Optical_xy2}, the absorptive parts of optical conductivity, $\sigma_{xx}^1$ and $\sigma_{xy}^2$, have direct physical interpretations, which measures the average and difference in absorptions of the left- and right-circularly polarized light, respectively.  The $\sigma_{xx}^1$ plotted in Fig.~\ref{fig:optical}(a) exhibits two sharp absorption peaks at 0.6 and 2.2 eV.  Since $\sigma_{xx}^1$ is directly related to the interband transition probability and jointed density of states, it is not affected by the spin direction, similarly to Mn$_3X$N ($X$ = Ga, Zn, Ag, or Ni)~\cite{Zhou2019}.  On the other hand, the $\sigma_{xy}^2$ plotted in Fig.~\ref{fig:optical}(d) oscillates drastically in the low-energy region and tends to zero above 5.5 eV.  The positive and negative values of $\sigma_{xy}^2$ indicate that the interband transitions are dominated by the excitations caused by the left- and right-circularly polarized light, respectively.  The signs of $\sigma_{xy}^2$ for the states of $S(90^\circ, 30^\circ)$ and $S(90^\circ, 90^\circ)$ are opposite, which has the same physical mechanism of the intrinsic AHC for monolayer LaCl~\cite{Liu2018}.  
%That is, starting from the unit cell of $\varphi=30^\circ$, an anticlockwise rotation of $60^\circ$ followed by an inversion operation restores the crystal structure but reverses the spin direction, resulting in the sign change of $\sigma_{xy}^2$.
It should be further noticed that for the state of $S(90^\circ, 0^\circ)$, $\sigma_{xy}^2$ is suppressed due to the presence of the mirror plane $\mathcal{M}$ that is perpendicular to $S$, which is consistent with the previous group theory analyses.  Utilizing the Kramers-Kronig transformation, the dispersive parts of optical conductivity, $\sigma_{xx}^2$ and $\sigma_{xy}^1$, can be obtained from the corresponding absorptive parts according to Eqs.~\eqref{eq:Optical_xx2} and~\eqref{eq:Optical_xy1}.  The dependence of $\sigma_{xx}^2$ and $\sigma_{xy}^1$ on the magnetization direction, featured in Figs.~\ref{fig:optical}(b) and~\ref{fig:optical}(c), resemble that of $\sigma_{xx}^1$ and $\sigma_{xy}^2$.

Then, we proceed to the out-of-plane magnetization by considering the spin within the $yz$ plane, for example $S(0^\circ, 90^\circ)$ ($S\parallel z$) and $S(45^\circ, 90^\circ)$.  As shown in Figs.~\ref{fig:optical}(e) and~\ref{fig:optical}(f), $\sigma_{xx}^1$ has two absorption peaks at 0.6 and 2.2 eV, and meanwhile $\sigma_{xx}^2$ presents two valleys at 0.5 and 2.0 eV.  This is identical to the situation of in-plane magnetization and further indicates that the diagonal elements of optical conductivity are not affected by the spin direction.  In contrast, the off-diagonal elements of optical conductivity, $\sigma_{xy}^1$ and $\sigma_{xy}^2$ [see Figs.~\ref{fig:optical}(g) and~\ref{fig:optical}(h)], obviously depend on the spin direction.  $\sigma_{xy}^1$ and $\sigma_{xy}^2$ oscillate as a function of photon energy with different spin directions and reach the maximal values when the spin points towards the $z$-axis.  It is important to notice that the off-diagonal elements of optical conductivity with the out-of-plane magnetization are enhanced by about one order of magnitude compared to that of in-plane magnetization.

\begin{figure*}[htbp]
	\includegraphics[width=2\columnwidth]{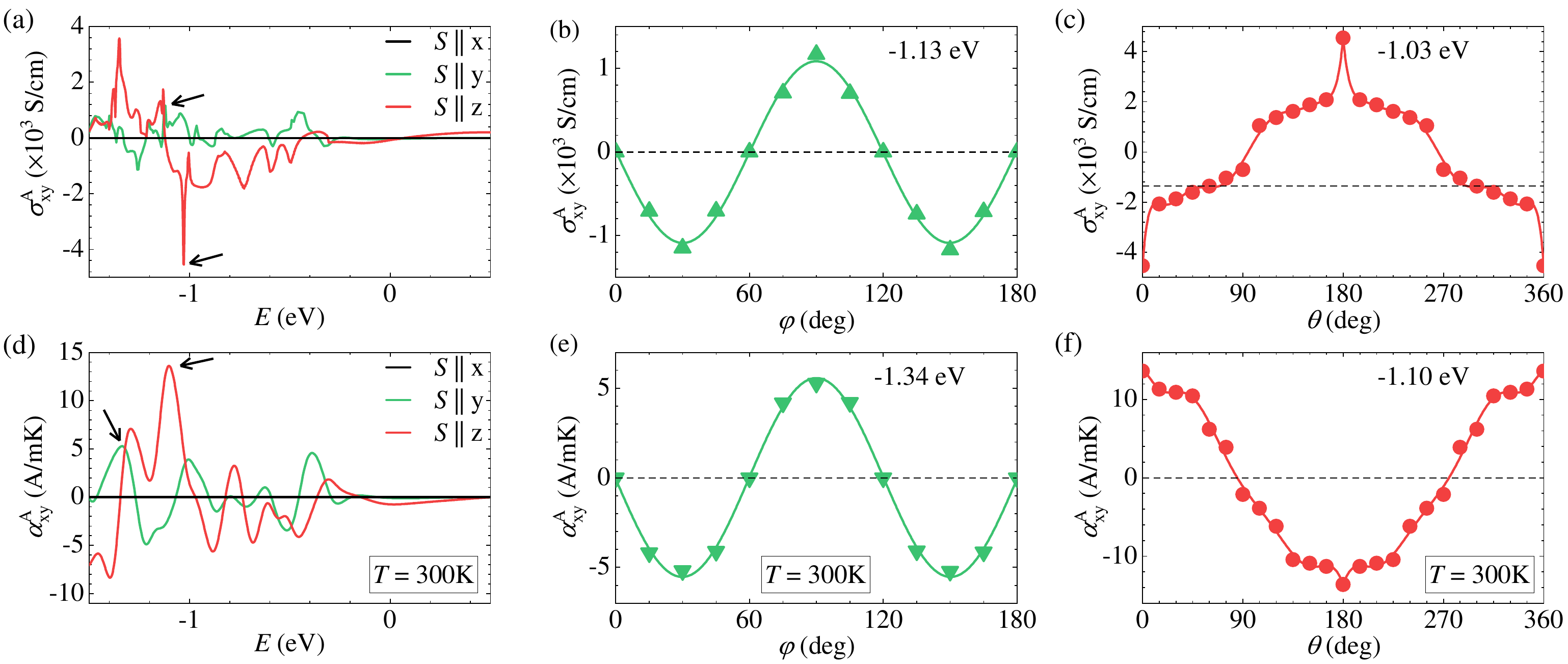}
	\caption{(Color online) (a,d) The intrinsic anomalous Hall ($\sigma^\textnormal{A}_{xy}$) and anomalous Nernst ($\alpha^\textnormal{A}_{xy}$) conductivity for monolayer $1T$-CrTe$_2$ as a function of the Fermi energy when the spin points along the $x$-, $y$-, and $z$-axis.  $\alpha^\textnormal{A}_{xy}$ is calculated at the temperature of 300 K.  The black arrows indicates the maximal values of $\sigma^\textnormal{A}_{xy}$ and $\alpha^\textnormal{A}_{xy}$.  (b,e) $\sigma^\textnormal{A}_{xy}$ and $\alpha^\textnormal{A}_{xy}$ as a function of azimuthal angle $\varphi$ when the Fermi energies are set to be -1.13 and -1.34 eV, respectively.  (c,f) $\sigma^\textnormal{A}_{xy}$ and $\alpha^\textnormal{A}_{xy}$ as a function of polar angle $\theta$ when the Fermi energies are set to be -1.03 and -1.10 eV, respectively.}
	\label{fig:ahc_anc}
\end{figure*}

\begin{figure*}
	\includegraphics[width=1.6\columnwidth]{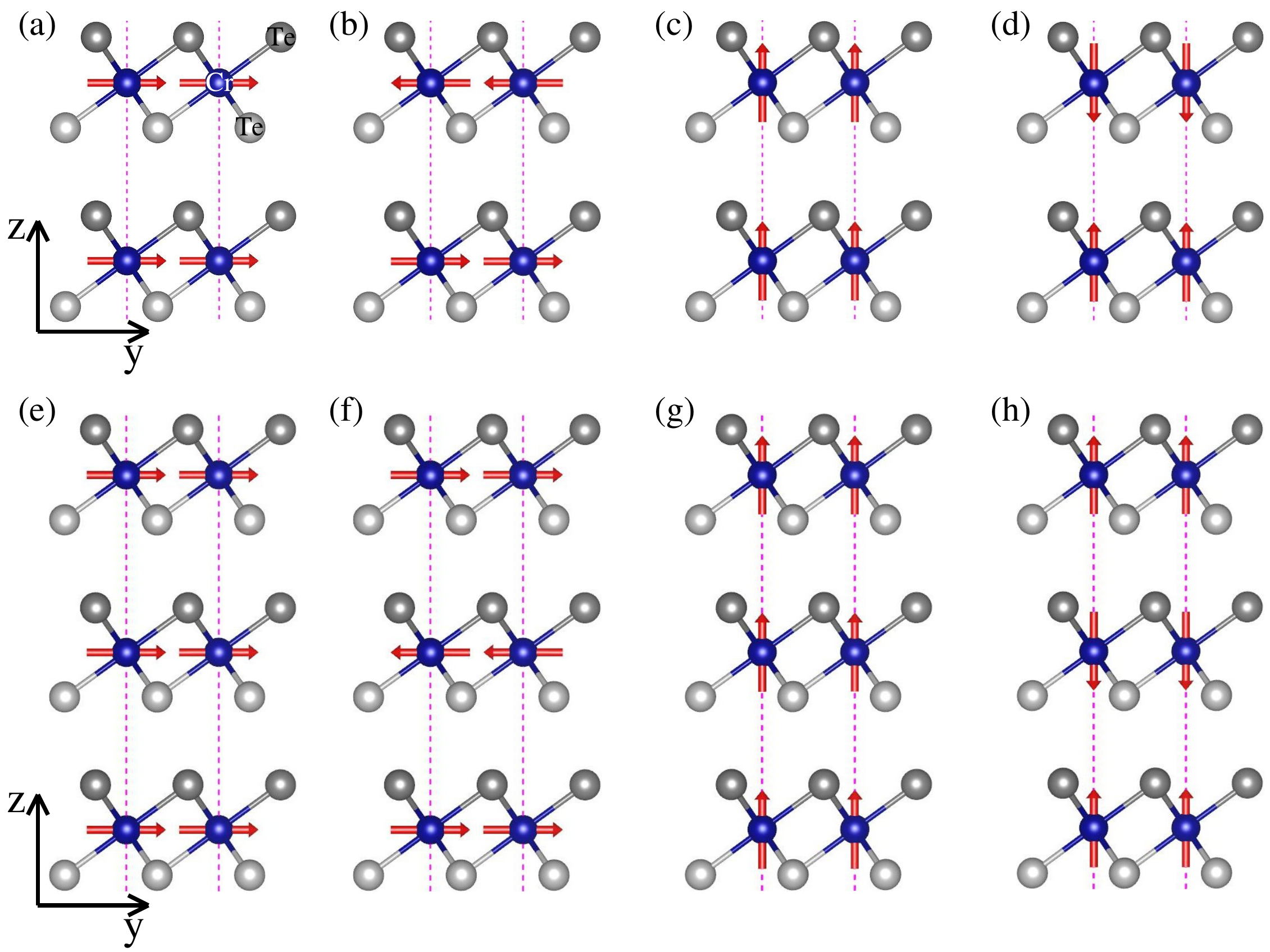}
	\caption{(Color online) Side views of bilayer (a-d) and trilayer (e-h) $1T$-CrTe$_2$ with in-plane and out-of-plane ferromagnetic and antiferromagnetic configurations.  The red arrows label the spin directions.}
	\label{fig:crystal_ML}
\end{figure*}

Now, we present the magneto-optical Kerr and Faraday spectra with in-plane and out-of-plane magnetization, as shown in Figs.~\ref{fig:MOE1} and~\ref{fig:MOE2}, respectively.  The Kerr and Faraday spectra are rather similar to that of $\sigma_{xy}$, and the reason can be simply attributed to their close relationships [refer to Eqs.~\eqref{eq:Kerr} and~\eqref{eq:Faraday2}].  For the in-plane magnetization, the Kerr and Faraday angles are vanishing if the spin points along $\varphi= n\pi/3$, for example $\varphi=0^\circ$ [see Figs.~\ref{fig:MOE1}(a-d)], due to the symmetry restriction.  When the spin rotates to $\varphi = \pi/6 + n\pi/3$, the Kerr and Faraday angles reach their maximums, that is, $\theta_\textnormal{K}^{\textnormal{max}}=0.24$ deg and $\theta_\textnormal{F}^{\textnormal{max}}=3.00\times10^5$ deg/cm at the photon energies of 0.70 and 0.79 eV, respectively.  The $\theta_\textnormal{K}^{\textnormal{max}}$ of CrTe$_2$ is comparable with the Kerr rotation angles of monolayer CrI$_3$ (0.286 deg)~\cite{Huang2017} and of blue phosphorene (0.12 deg)~\cite{Zhou2017}.  Moreover, Figs.~\ref{fig:MOE1}(e-h) show that the maximal Kerr and Faraday angles of monolayer CrTe$_2$ exhibits a period of $2\pi/3$ when the spin rotates within the $xy$ plane.

On the other hand, the Kerr and Faraday spectra with the out-of-plane magnetization are illustrated in Fig.~\ref{fig:MOE2}.  Inheriting from the off-diagonal elements of optical conductivity, the Kerr and Faraday spectra with out-of-plane magnetization are significantly stronger than that with in-plane magnetization [Figs.~\ref{fig:MOE2}(a-d)].  The maximal Kerr and Faraday rotation angles appear when $\theta=0^\circ$ ($S\parallel z$), that is, $\theta_\textnormal{K}^{\textnormal{max}} = -1.34$ deg and $\theta_\textnormal{F}^{\textnormal{max}} = -17.30\times10^5$ deg/cm at the photon energies of 0.66 and 0.82 eV, respectively.  Moreover, the Kerr and Faraday angles have a period of $2\pi$ as a function of polar angle $\theta$, as shown in Figs.~\ref{fig:MOE2}(e-h), demonstrating again that the MOE can be effectively modulated by tuning the spin direction.

\begin{figure}
	\includegraphics[width=0.9\columnwidth]{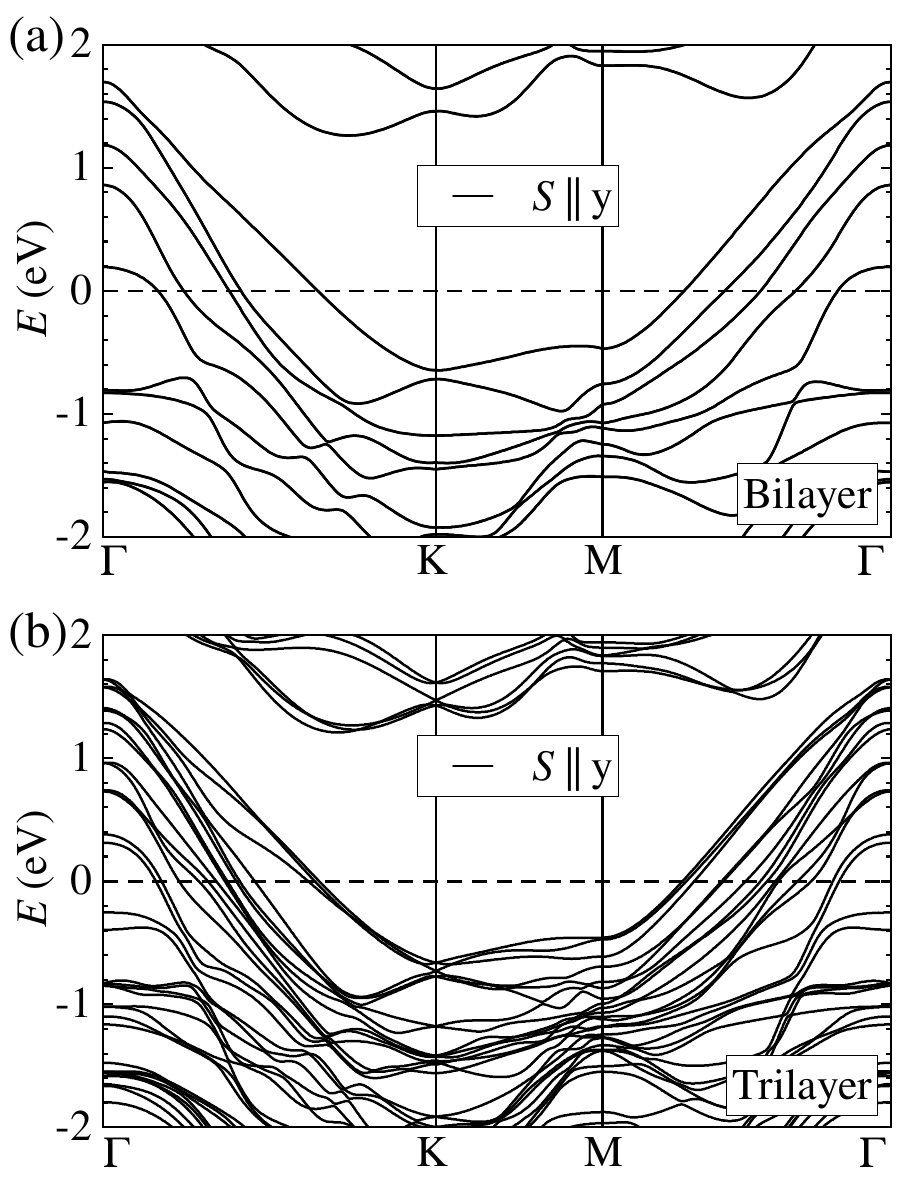}
	\caption{(Color online) Relativistic band structures of bilayer and trilayer $1T$-CrTe$_2$ when the spin is along the $y$-axis.}
	\label{fig:band_ML}
\end{figure}

\subsubsection{Anomalous Hall and anomalous Nernst effects}

As mentioned above, the dc limit of real off-diagonal element of optical conductivity, i.e., $\sigma_{xy}^1(\omega \rightarrow 0)$, is nothing but the AHC ($\sigma^\textnormal{A}_{xy}$), which can be alternatively evaluated by integrating the Berry curvature over the entire Brillouin zone [see Eq.~\eqref{eq:AHC}]~\cite{Yao2004}.  Figure~\ref{fig:ahc_anc}(a) plots the AHC as a function of the Fermi energy when the spin points along the $x$-, $y$-, and $z$-axis.  $\sigma^\textnormal{A}_{xy}$ is vanishing when $S\parallel x$ and turns to appear if $S\parallel y$ and $z$, indicating the same symmetry requirements for the MOE.  In analogy to the Kerr and Faraday angles, the $\sigma^\textnormal{A}_{xy}$ with out-of-plane magnetization ($S\parallel z$) is significantly larger than that with in-plane magnetization ($S\parallel y$).  Thus, at the actual Fermi energy, the $\sigma^\textnormal{A}_{xy}$ with both in-plane and out-of-plane magnetization are relatively small, which is adverse to measure experimentally.  Nevertheless, the pronounced peaks of AHC arise after appropriate holes are introduced.  For example, $\sigma^\textnormal{A}_{xy}$ can increase up to -4539.23 S/cm at -1.03 eV when $S\parallel z$ and up to 1168.12 S/cm at -1.13 eV when $S\parallel y$, respectively.  When the spin rotates within the $xy$ and $yz$ planes, $\sigma^\textnormal{A}_{xy}$ exhibits the periods of $2\pi/3$ and $2\pi$, respectively [depicted in Figs.~\ref{fig:ahc_anc}(b) and~\ref{fig:ahc_anc}(c)], which are identical to the behaviors of Kerr and Faraday angles.

The ANE, being regarded as the thermoelectric counterpart of the AHE, is a celebrated effect from the realm of spin caloritronics~\cite{Bauer2012,Boona2014}.  The conclusions of symmetry analyses for the AHE are also applicable to the ANE, comparing with Eqs.~\eqref{eq:AHC} and~\eqref{eq:ANC}.  That is, the ANC $\alpha^\textnormal{A}_{xy}$ is forbidden when the spin is along $\varphi= n\pi/3$, e.g., $S\parallel x$, and turns to be nonzero if $S\parallel y$ or $z$, as clearly shown in Fig.~\ref{fig:ahc_anc}(d).  Due to the high Curie temperature of CrTe$_2$ ($T_C>$ 300 K)~\cite{Freitas2015,Sun2020,Purbawati2020}, the first-principles calculations of the ANC are carried out at the room-temperature of 300 K.  Similarly to the AHC $\sigma^\textnormal{A}_{xy}$, the ANC $\alpha^\textnormal{A}_{xy}$ with out-of-plane magnetization ($S\parallel z$) is evidently larger than that with in-plane magnetization ($S\parallel y$).  For both in-plane and out-of-plane magnetization, $\alpha^\textnormal{A}_{xy}$ are almost zero at the actual Fermi energy and give rise to pronounced peaks by hole doping.  For example, $\alpha^\textnormal{A}_{xy}$ reaches up to 13.61 A/mK at -1.10 eV when $S\parallel z$ and up to 5.30 A/mK at -1.34 eV when $S\parallel y$, respectively.  Moreover, $\alpha^\textnormal{A}_{xy}$ displays a period of $2\pi/3$ ($2\pi$) when the spin rotates within the $xy$ ($yz$) plane, as shown in Figs.~\ref{fig:ahc_anc}(e) and~\ref{fig:ahc_anc}(f), just like the AHC as well the Kerr and Faraday angles.

\begin{table}[b]\footnotesize
	\caption{The total energy $E_{\textnormal{tot}}$ (in the unit of eV) per unit cell of bilayer (BL) and trilayer (TL) $1T$-CrTe$_2$ with in-plane and out-of-plane ferromagnetic (i-FM and o-FM) and antiferromagnetic (i-AFM and o-AFM) configurations.  The superscripts $a-h$ corresponds to the magnetic structures presented in Fig.~\ref{fig:crystal_ML}.  The energies of nonmagnetic (NM) structures are given for the reference.  The relaxed lattice constant $a$ (in the unit of \AA) is also listed.}
	\label{tab:energy}
	\begin{ruledtabular}
		\begingroup
		\setlength{\tabcolsep}{4.5pt}
		\renewcommand{\arraystretch}{1.5}
		\begin{tabular}{lrrrrrr}
			&&   i-FM$^{a,e}$ &    o-FM$^{c,g}$ &   i-AFM$^{b,f}$ & o-AFM$^{d,h}$   & NM  \\
			\hline
			BL&$E_{\textnormal{tot}}$   & -31.996 & -31.983   & -32.021   & -32.017  & -28.885  \\
			&$a$     & 3.763   & 3.754     & 3.784     & 3.786    & 3.469    \\
			%\hline
			TL &$E_{\textnormal{tot}}$   & -48.216 & -48.208   & -48.238   & -48.237  & -43.528  \\
			&$a$    & 3.778   & 3.778     & 3.791     & 3.793    & 3.478    \\
		\end{tabular}
		\endgroup
	\end{ruledtabular}
\end{table}

\begin{figure*}
	\includegraphics[width=1.6\columnwidth]{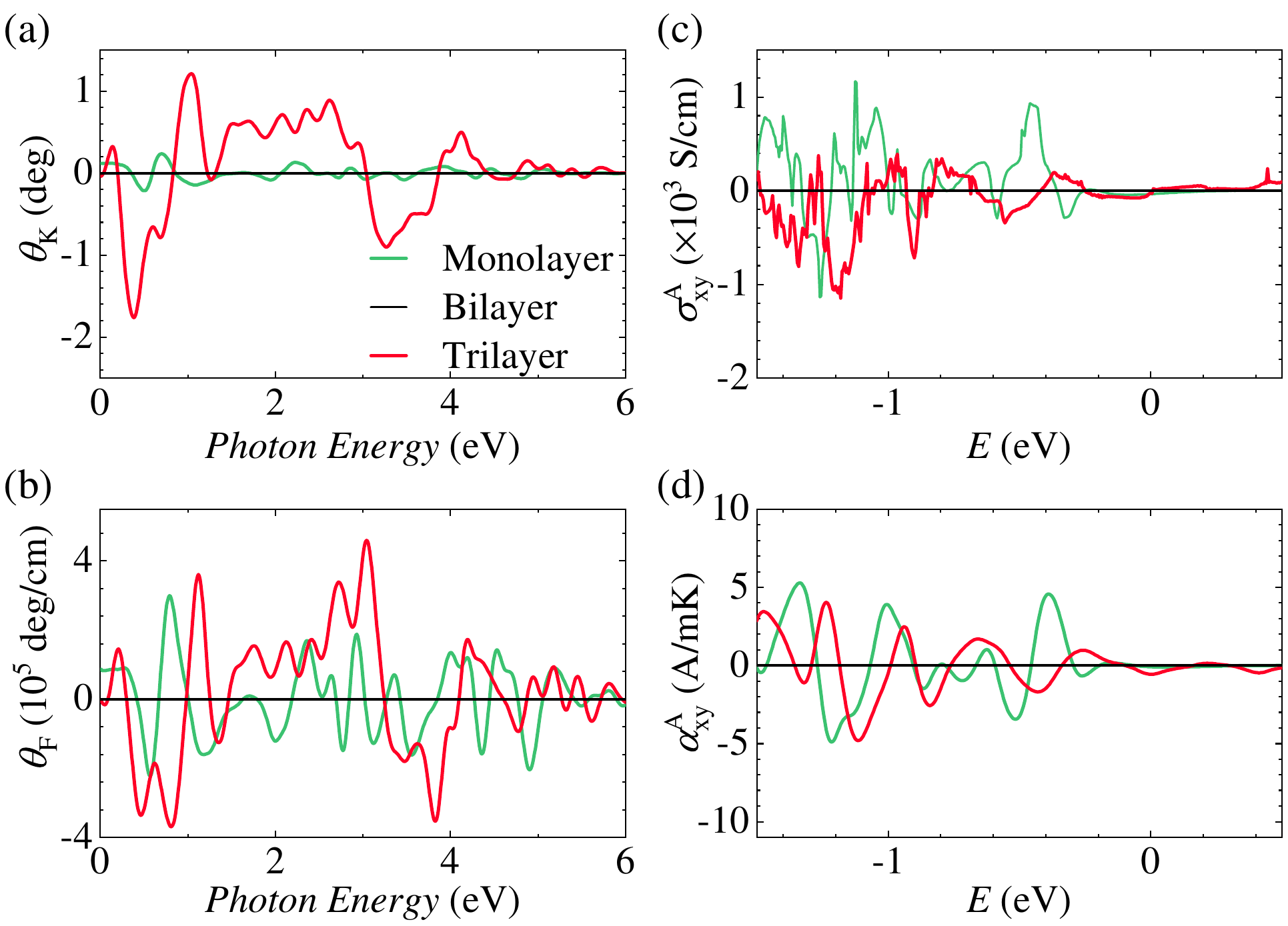}
	\caption{(Color online) The Kerr (a) and Faraday (b) rotation angles of bilayer and trilayer $1T$-CrTe$_2$.  The anomalous Hall and anomalous Nernst conductivities of bilayer and trilayer $1T$-CrTe$_2$ as a function of the Fermi energy.  The magnetization direction is along the $y$-axis.  For a better comparison, the curves of monolayer $1T$-CrTe$_2$ are also plotted.}
	\label{fig:multilayer}
\end{figure*}

\subsection{Multilayer CrTe$_2$}
In this subsection, we shall discuss the layer-dependent magnetic properties as well as the MOE, AHE, and ANE of multilayer CrTe$_2$.  The bilayer and trilayer CrTe$_2$ with the AA-stacking pattern, which could be directly exfoliated from the bulk structure, are considered here.  To determine the magnetic ground states, we calculate the total energy ($E_{\textnormal{tot}}$) of in-plane and out-of-plane ferromagnetic and antiferromagnetic structures, as depicted in Fig.~\ref{fig:crystal_ML}.  It should be stressed here that all the trilayer CrTe$_2$ are actually ferromagnetic with finite net magnetization, while we mention the ``antiferromagnetic" trilayer structures [Figs.~\ref{fig:crystal_ML}(f,h)] just because of the interlayer antiferromagnetic order.  The energy results are summarized in Tab.~\ref{tab:energy}, from which one can find that the antiferromagnetic structures for both bilayer and trilayer CrTe$_2$ [Figs.~\ref{fig:crystal_ML}(b,d) and~\ref{fig:crystal_ML}(f,h)] are energetically favorable.  Thus, the in-plane antiferromagnetic structures  [Figs.~\ref{fig:crystal_ML}(b) and~\ref{fig:crystal_ML}(f)] are most stable with the slightly lower energies of $\sim$4 and $\sim$1 meV/cell than the out-of-plane ones for bilayer and trilayer, respectively.  In the following, we only focus on the  bilayer and trilayer CrTe$_2$ with the in-plane antiferromagnetic structures.  The electronic band structures plotted in Fig.~\ref{fig:band_ML} demonstrate the metallic nature of both bilayer and trilayer CrTe$_2$.

Using the group theory, we analyze whether the MOE, AHE, and ANE can exist in bilayer and trilayer CrTe$_2$.  The magnetic point group of in-plane antiferromagnetic bilayer [Fig.~\ref{fig:crystal_ML}(b)] is $2/m'$, which contains the spacetime inversion symmetry $\mathcal{TP}$ that forbids any signals of magneto-optical responses as well as anomalous charge and thermoelectric transports.  In contrast, the magnetic point group of in-plane antiferromagnetic trilayer [Fig.~\ref{fig:crystal_ML}(f)] is the same as that of monolayer CrTe$_2$, i.e., $2'/m'$, which allows the presence of the MOE, AHE, and ANE.

The layer number can influence the MOE, AHE, and ANE of multilayer CrTe$_2$.  The magneto-optical Kerr and Faraday rotation angles of in-plane antiferromagnetic bilayer and trilayer CrTe$_2$ are plotted in Figs.~\ref{fig:multilayer}(a) and~\ref{fig:multilayer}(b), in which the results of monolayer CrTe$_2$ are given for comparison.  As expected, the $\theta_\textnormal{K}$ and $\theta_\textnormal{F}$ of bilayer structure are zero due to the presence of $\mathcal{TP}$ symmetry.  For the trilayer structure, the largest $\theta_\textnormal{K}$ and $\theta_\textnormal{F}$ are -1.76 deg and 4.60$\times 10^5$ deg/cm at the photon energies of 0.38 and 3.04 eV, respectively.  One can find that the Kerr and Faraday effects of trilayer structure are generally stronger than that of monolayer structure.  Moreover, the AHC and ANC of monolayer, bilayer, and trilayer CrTe$_2$ with in-plane magnetization are presented in Figs.~\ref{fig:multilayer}(c) and~\ref{fig:multilayer}(d), respectively.  Clearly, the $\sigma^{\textnormal{A}}_{xy}$ and $\alpha^{\textnormal{A}}_{xy}$ of bilayer structure are vanishing due to the symmetry restriction.  Although the $\sigma^{\textnormal{A}}_{xy}$ and $\alpha^{\textnormal{A}}_{xy}$ of  trilayer structure are very small at the actual Fermi energy, they can be significantly enhanced by hole doping.  The largest $\sigma^{\textnormal{A}}_{xy}$ and $\alpha^{\textnormal{A}}_{xy}$ are 1147.03 S/cm at -1.18 eV and -4.81 A/mK at -1.12 eV, respectively.  The AHE and ANE of trilayer structure are slightly smaller than that of monolayer structure.

\section{Summary}
In summary, using the first-principles density functional theory calculations and group theory analyses, we have systematically investigated the electronic, magnetic, magneto-optical, anomalous charge and thermoelectric transport properties of monolayer, bilayer, and trilayer $1T$-CrTe$_2$.  The monolayer is a ferromagnetic metal with the in-plane magnetization along the $y$-axis.  The in-plane magnetocrystalline anisotropy energy is as small as 10 $\mu$eV/cell, indicating that the spin can be easily rotated within the $xy$ plane.  The magneto-optical Kerr and Faraday rotation angles as well as anomalous Hall and Nernst conductivities exhibit a period of $2\pi/3$ when the spin rotates within the $xy$ plane, and their maximums of $\theta_\textnormal{K}$ = 0.24 deg, $\theta_\textnormal{F}$ = 3.00$\times10^5$ deg/cm, $\sigma^\textnormal{A}_{xy}$ = 1168.12 S/cm, and $\alpha^\textnormal{A}_{xy}$ = 5.30 A/mK (300 K) appear at $\varphi = n\pi/3 + \pi/6$ with $n \in \mathbb{N}$. At the azimuthal angle $\varphi = n\pi/3$, the mirror planes that are normal to the spin direction suppress the magneto-optical, anomalous Hall, and anomalous Nernst effects.  If the spin cants from in-plane to out-of-plane direction, the magneto-optical, anomalous Hall, and anomalous Nernst effects are significantly enhanced, and particularly they reach to the maximal values of $\theta_\textnormal{K}$ = -1.34 deg, $\theta_\textnormal{F}$ = -17.30$\times10^5$ deg/cm, $\sigma^\textnormal{A}_{xy}$ = -4539.23 S/cm, and $\alpha^\textnormal{A}_{xy}$ = 13.61 A/mK (300 K) when the spin is along the $z$-axis (i.e., polar angle $\theta=0^\circ$).  The bilayer $1T$-CrTe$_2$ prefers an in-plane antiferromagnetic structure with the magnetization along the $y$-axis, which has the spacetime inversion symmetry $\mathcal{TP}$ that prohibits the signals of magneto-optical responses as well as anomalous Hall and Nernst transports.  The trilayer $1T$-CrTe$_2$ is also inclined to the in-plane antiferromagnetic order between two adjacent layers, but has finite net magnetization due to the odd number of layers.  The magnetic point group of trilayer structure with in-plane antiferromagnetic order is identical to that of monolayer structure and thus allows the presence of all the physical phenomena mentioned above.  In particular, the magneto-optical Kerr and Faraday rotation angles (anomalous Hall and Nernst conductivities) of trilayer structure are obviously larger (slightly smaller) than that of monolayer structure with the magnetization along the $y$-axis.  For example, the maximal values of $\theta_\textnormal{K}$ = -1.76 deg, $\theta_\textnormal{F}$ = 4.60$\times10^5$ deg/cm, $\sigma^\textnormal{A}_{xy}$ = -1147.03 S/cm, and $\alpha^\textnormal{A}_{xy}$ = -4.81 A/mK (300 K) are found in the trilayer structure .  Our results suggest that the magneto-optical, anomalous Hall, and anomalous Nernst effects for two-dimensional room-temperature ferromagnet $1T$-CrTe$_2$ can be effectively modulated by altering magnetization direction and layer number.

\begin{acknowledgments}
W.F. and Y.Y. acknowledge the support from the National Natural Science Foundation of China (Grants No. 11874085 and No. 11734003) and the National Key R\&D Program of China (Grant No. 2016YFA0300600).  X.Z. acknowledges the support from the Graduate Technological Innovation Project of Beijing Institute of Technology (Grant No. 2019CX10018).  
\end{acknowledgments}
	
%\bibliography{references}

%\end{document}


\end{document}